\documentclass[twocolumn]{aastex631}  

\usepackage{newtxtext,newtxmath}
\usepackage[T1]{fontenc}

\usepackage{nicefrac}
\usepackage{amssymb}
\usepackage{latexsym}
\usepackage{amsmath, mathrsfs, bm}
\usepackage{eqnarray}
\usepackage{url}
\usepackage{cases}
\usepackage{epsfig}
\usepackage{color}
\usepackage{graphicx}
\usepackage{grffile}
\usepackage{float}
\usepackage{soul}
\usepackage{enumerate}
\usepackage{multirow}
\usepackage{threeparttable}
\usepackage{xspace}

\usepackage{bm}
\usepackage{times}

\usepackage{etoolbox}
\usepackage{multirow}

\begin{document}

\title{Resolving the Hubble Tension in the Early Dark Energy Framework with JWST and DESI Data}

\author[0009-0005-6921-3201]{Guo-Hong Du} 
\affiliation{Liaoning Key Laboratory of Cosmology and Astrophysics, College of Sciences, Northeastern University, Shenyang 110819, China}

\author[0009-0004-6982-4021]{Tian-Nuo Li}\thanks{} 
\affiliation{Liaoning Key Laboratory of Cosmology and Astrophysics, College of Sciences, Northeastern University, Shenyang 110819, China}

\author[0000-0002-8289-0814]{Lu Yin}  
\affiliation{Department of Physics, Shanghai University, Shanghai 200444,  China}  

\author[0009-0003-6056-5257]{Sheng-Han Zhou} 
\affiliation{Liaoning Key Laboratory of Cosmology and Astrophysics, College of Sciences, Northeastern University, Shenyang 110819, China}

\author[0000-0003-0232-8344]{Hao Wang} 
\affiliation{School of Fundamental Physics and Mathematical Sciences, Hangzhou Institute for Advanced Study, UCAS, Hangzhou 310024, China}
\affiliation{School of Physical Sciences, University of Chinese Academy of Sciences, Beijing 100049, China}

\author[0000-0002-3512-2804]{Jing-Fei Zhang} 
\affiliation{Liaoning Key Laboratory of Cosmology and Astrophysics, College of Sciences, Northeastern University, Shenyang 110819, China}

\author[0000-0002-6029-1933]{Xin Zhang}\thanks{} 
\affiliation{Liaoning Key Laboratory of Cosmology and Astrophysics, College of Sciences, Northeastern University, Shenyang 110819, China}
\affiliation{MOE Key Laboratory of Data Analytics and Optimization for Smart Industry, Northeastern University, Shenyang 110819, China}
\affiliation{National Frontiers Science Center for Industrial Intelligence and Systems Optimization, Northeastern University, Shenyang 110819, China}

\correspondingauthor{Xin Zhang}
\email{zhangxin@mail.neu.edu.cn}
\correspondingauthor{Tian-Nuo Li}
\email{litiannuo@stumail.neu.edu.cn}

\begin{abstract}
In the JWST and DESI era, the JWST high-redshift galaxy observations and DESI baryon acoustic oscillation (BAO) measurements severely challenge the standard $\Lambda$CDM model, while the $H_0$ tension becomes increasingly prominent. In this work, we investigate the capability of the early dark energy (EDE) model to alleviate the $H_0$ tension utilizing cosmic microwave background data from Planck, ACT, and SPT, BAO data from DESI, and ultraviolet luminosity function observations from the JWST. Within the canonical axion EDE framework, the CMB+DESI+JWST data significantly increase the $H_0$ value to $71.58\pm1.05\,\mathrm{km\,s^{-1}\,Mpc^{-1}}$ with the residual $H_0$ tension quantified by $Q_{\rm DMAP}$ reduced to $0.6\sigma$. Simultaneously, this model improves the fit to the JWST data and exhibits statistical performance significantly better than the $\Lambda$CDM model, with $\Delta\chi^2_{\mathrm{tot}} = -18.26$ and $\Delta\mathrm{DIC} = -11.89$. Our results highlight the complementary advantages of JWST high-redshift galaxy data alongside early- and late-time observations in testing EDE and alleviating the $H_0$ tension.
\end{abstract}

\section{Introduction}

The Hubble constant $H_0$ represents the present expansion rate and anchors the cosmic distance scale. Within $\Lambda$CDM, the final Planck analysis infers $H_0=67.4\pm0.5~\rm km\,s^{-1}\,Mpc^{-1}$ from cosmic microwave background (CMB) observables~\citep{Planck:2018vyg}, while the local distance ladder calibrated with Cepheids gives $H_0=73.04\pm1.04~\rm km\,s^{-1}\,Mpc^{-1}$ from SH0ES~\citep{Riess:2021jrx}. This $5.3\sigma$ discrepancy has persisted through many checks of observational systematics and constitutes one of the most prominent tensions in precision cosmology~\citep{Zhao:2017urm,Verde:2019ivm,Guo:2018ans,Vagnozzi:2019ezj,DiValentino:2020zio,DiValentino:2021izs,Shah:2021onj,Vagnozzi:2021gjh,Gao:2021xnk,Perivolaropoulos:2021jda,Schoneberg:2021qvd,Abdalla:2022yfr,DiValentino:2022fjm,Kamionkowski:2022pkx,Giare:2023xoc,Hu:2023jqc,Vagnozzi:2023nrq,Bernui:2023byc,Jiang:2024xnu,Giare:2024akf,Gao:2022ahg,Jiang:2025ylr,Pedrotti:2024kpn,Pedrotti:2025ccw,Pedrotti:2026dwj}. JWST observations of Cepheids in SN hosts provide an independent high-resolution test of Hubble Space Telescope crowding and blending effects and remain consistent with the local distance-ladder calibration, with representative combined values near $H_0=73.17\pm0.86~\rm km\,s^{-1}\,Mpc^{-1}$~\citep{Riess:2024ohe}. If both the early- and late-Universe determinations are correct, this tension may point to missing physics in the standard cosmological model or to a modification of the assumptions connecting early-Universe observables to the present expansion rate~\citep{Boisseau:2000pr,Chevallier:2000qy,Li:2004rb,Huang:2004wt,Zhang:2005hs,Zhang:2005yz,Zhang:2007sh,Zhang:2009un,Zhang:2014nta,Zhang:2015rha,Cai:2015emx,Wang:2016lxa,Feng:2016djj,Guo:2015gpa,Nojiri:2017ncd,Guo:2018gyo,Feng:2019jqa,Yin:2023srb,Jin:2022qnj,Yao:2023qve,Song:2022siz,Wang:2024vmw,Giare:2024smz,CosmoVerseNetwork:2025alb,Kochappan:2024jyf}.

The connection between the $H_0$ tension and early-Universe physics can be understood from the CMB acoustic scale. The CMB tightly measures
\begin{equation}
\theta_{\rm s}=\frac{r_{\rm s}(z_*)}{D_{\rm A}(z_*)},
\end{equation}
where $z_*$ is the photon-decoupling redshift and $D_{\rm A}(z_*)$ is the angular diameter distance to last scattering. The comoving sound horizon is
\begin{equation}
r_{\rm s}(z_*)=\int_{z_*}^{\infty}\frac{c_{\rm s}(z)}{H(z)}\,{\rm d}z,
\end{equation}
where $c_{\rm s}$ is the sound speed of the photon-baryon fluid. Since $\theta_{\rm s}$ is measured with very high precision, increasing the pre-recombination expansion rate reduces $r_{\rm s}$ and can be compensated by a smaller $D_{\rm A}(z_*)$, which in spatially flat models is usually associated with a larger inferred $H_0$. This provides the basic motivation for early-time solutions to the $H_0$ tension: they raise the CMB-inferred $H_0$ by reducing the sound horizon while keeping the angular acoustic scale approximately unchanged~\citep{Knox:2019rjx}.

Early dark energy (EDE) realizes this mechanism by introducing a transient energy component that contributes non-negligibly before or around recombination and then rapidly dilutes~\citep{Poulin:2018cxd,Lin:2019qug,Ye:2020btb,Ye:2020oix,Braglia:2020bym,Agrawal:2019lmo,Poulin:2023lkg,Bella:2026zuk,Yin:2026gss}. In its simplest form, EDE behaves approximately as vacuum energy while the scalar field is frozen by Hubble friction, becomes dynamical near a critical redshift $z_{\rm c}$, and subsequently redshifts away faster than matter. The canonical axion-like EDE scenario demonstrated that this mechanism can reduce the sound horizon and raise the CMB-inferred $H_0$~\citep{Poulin:2018cxd}. However, it is tightly constrained by the CMB damping tail, CMB lensing, baryon acoustic oscillation (BAO), SN, and large-scale structure data~\citep{Hill:2020osr,Ivanov:2020ril,Smith:2019ihp}. Several theoretically distinct realizations have therefore been proposed, including Rock `n' Roll EDE based on power-law scalar dynamics~\citep{Agrawal:2019lmo}, $\alpha$-attractor EDE motivated by plateau-like scalar potentials~\citep{Braglia:2020bym}, and the AdS/axion-inspired EDE model~\citep{Ye:2020btb}. These models share the same sound-horizon motivation but differ in their potentials, perturbation dynamics, post-transition equation of state, and phenomenological viability.

With increasing precision and statistical power of current cosmological observations, tests of EDE have entered a new stage. CMB observations remain the primary probe of its dynamics, while Planck provides the baseline measurements and recent ACT and SPT observations add high-resolution temperature and polarization information on small angular scales. At lower redshifts, DESI DR2 BAO measurements test the consistency between the early sound horizon and the late-time expansion history through $D_{\mathrm{M}}/r_{\mathrm{d}}$, $D_{\mathrm{H}}/r_{\mathrm{d}}$, and $D_{\mathrm{V}}/r_{\mathrm{d}}$~\citep{DESI:2025zgx}. The DESI BAO results have already motivated a broad range of cosmological analyses beyond $\Lambda$CDM~\citep{Giare:2024gpk,Jiang:2024xnu,Yang:2024kdo,Li:2024qso,Ye:2024ywg,Reboucas:2024smm,Park:2024pew,Li:2024qus,Wolf:2025jed,Shajib:2025tpd,Giare:2025pzu,Chaussidon:2025npr,Pang:2025lvh,RoyChoudhury:2025dhe,Paliathanasis:2025cuc,Scherer:2025esj,Giare:2024oil,Liu:2025mub,Teixeira:2025czm,Cheng:2025lod,Cai:2025mas,Li:2025ops,Du:2025xes,Du:2024pai,Wang:2024dka,Li:2025vuh,Specogna:2025guo,Song:2025bio,Li:2026xaz,Du:2025csv,Du:2026cly,Li:2025owk,Li:2026asg,Sabogal:2026qvy,Sabogal:2026ipu}. When combined with SN samples such as DESY5 and PantheonPlus, BAO+SN data further constrain the late-time distance-redshift relation~\citep{Brout:2022vxf,DES:2024jxu}. Thus, any EDE-induced reduction of the sound horizon must simultaneously satisfy CMB and low-redshift distance measurements.

Complementary to the CMB, BAO, and SN constraints, JWST provides new information on both the local distance ladder and high-redshift structure formation. JWST Cepheid observations directly test the calibration of the local distance ladder~\citep{Riess:2024ohe}, while massive and luminous galaxy candidates at $z\gtrsim7$--$12$ provide a complementary probe of early structure formation, subject to assumptions about stellar masses, halo mass functions, and galaxy formation efficiency. Since EDE modifies the early expansion history and may affect the growth and timing of structure formation, JWST galaxy data provide an independent test of whether an early component capable of reducing the sound horizon remains compatible with the observed abundance of early massive galaxies~\citep{Shen:2024hpx,Huang:2024aog,Forconi:2023hsj,Jiang:2024tll}.

In this letter, we use the latest CMB, DESI DR2 BAO, SN, and JWST data to systematically explore multiple EDE scenarios. We find that the axion-like EDE model, in particular, can alleviate the $H_0$ tension to the $0.6\sigma$ level while improving the fit to JWST high-redshift galaxy observations. Our results suggest that, in the DESI and JWST era, EDE remains a viable and testable pre-recombination extension of $\Lambda$CDM.

\section{Methodology} \label{sec2}

\subsection{EDE Models}

To investigate the physical implications of the EDE scenario, we consider four representative scalar-field models within a spatially flat Friedmann-Lema\^{i}tre-Robertson-Walker background. The total expansion rate is governed by the Friedmann equation
\begin{equation}
H^2=\frac{1}{3M_{\rm Pl}^2}\left(\rho_{\mathrm{r}}+\rho_{\mathrm{b}}+\rho_{\mathrm{c}}+\rho_\Lambda+\rho_\phi\right),
\end{equation}
where $M_{\rm Pl} = (8\pi G)^{-1/2}$ represents the reduced Planck mass with $G$ being the gravitational constant, while the subscripts $\mathrm{r}$, $\mathrm{b}$, $\mathrm{c}$, $\Lambda$, and $\phi$ denote radiation, baryons, cold dark matter, the cosmological constant, and the EDE component, respectively. For a homogeneous scalar field, $\rho_\phi=\frac{1}{2}\dot{\phi}^2+V(\phi)$ and $p_\phi=\frac{1}{2}\dot{\phi}^2-V(\phi)$, with its evolution determined by the Klein-Gordon equation
\begin{equation}
\ddot{\phi}+3H\dot{\phi}+\frac{{\rm d}V}{{\rm d}\phi}=0.
\end{equation}
The EDE contribution is characterized by its maximum fractional energy density $f_{\mathrm{EDE}}(z_{\mathrm{c}})\equiv\max_z[\rho_\phi(z)/\rho_{\mathrm{tot}}(z)]$ and the corresponding critical redshift $z_{\mathrm{c}}$.

The first realization is the Axion-EDE model~\citep{Poulin:2018cxd,Smith:2019ihp}, which adopts the periodic potential
\begin{equation}
V(\phi)=m^2f^2\left[1-\cos\left(\frac{\phi}{f}\right)\right]^n,
\end{equation}
where $m$ is the effective scalar-field mass, $f$ is the axion decay constant, and $n=3$ is fixed. The field is initially frozen by Hubble friction and becomes dynamical near $z_{\mathrm{c}}$, after which it oscillates with a cycle-averaged equation of state $\bar{w}=1/2$.

The second realization is the $\alpha$-attractor EDE ($\alpha$-EDE) model~\citep{Braglia:2020bym}, characterized by
\begin{equation}
V(\phi)=V_0\tanh^{2n}\left(\frac{\phi}{\sqrt{6\alpha}M_{\rm Pl}}\right),
\end{equation}
where $V_0$ is the potential amplitude, $\alpha$ determines the inverse curvature scale, and $n=2$ is fixed. Near the origin, the potential approaches a quartic form, giving a post-transition averaged equation of state $\bar{w}=1/3$.

The third realization is the Rock `n' Roll EDE (RnR-EDE) model~\citep{Agrawal:2019lmo}, which adopts the power-law potential
\begin{equation}
V(\phi)=V_0\left(\frac{\phi}{M_{\rm Pl}}\right)^{2n},
\end{equation}
with $n=2$. After remaining nearly frozen initially, the field oscillates around the potential minimum after $z_{\mathrm{c}}$, with a cycle-averaged equation of state $\bar{w}=1/3$.

The fourth realization is the anti-de Sitter EDE (AdS-EDE) model~\citep{Ye:2020btb}, which employs a piecewise potential with a negative AdS well,
\begin{equation}
V(\phi)=\begin{cases}V_0\left(\dfrac{\phi}{M_{\rm Pl}}\right)^4-V_{\mathrm{AdS}},&\text{for }\dfrac{\phi}{M_{\rm Pl}}<\left(\dfrac{V_{\mathrm{AdS}}}{V_0}\right)^{1/4},\\0,&\text{for }\dfrac{\phi}{M_{\rm Pl}}\ge\left(\dfrac{V_{\mathrm{AdS}}}{V_0}\right)^{1/4}.\end{cases}
\end{equation}
Here $V_0$ is the potential amplitude and $V_{\mathrm{AdS}}=\alpha_{\mathrm{AdS}}[\rho_{\mathrm{m}}(z_{\mathrm{c}})+\rho_{\mathrm{r}}(z_{\mathrm{c}})]$ parameterizes the depth of the negative potential well.

\subsection{JWST Observations}

We incorporate high-redshift galaxy observations from the JWST to independently constrain early structure formation. The observational quantity is the ultraviolet luminosity function, expressed as the logarithmic volume density $y_{\mathrm{obs}}\equiv\log_{10}\Phi$, measured over bins of the absolute ultraviolet magnitude $M_{\mathrm{UV}}$ in the redshift range $7\lesssim z\lesssim12$. Our baseline compilation combines the measurements of \citet{Perez-Gonzalez:2023wta,Casey:2024,Donnan:2024,Harikane:2025} with the faint-end extensions provided by \citet{Chemerynska:2026}.

{The theoretical ultraviolet luminosity function is modeled by mapping the underlying primordial matter perturbations to observable magnitudes,
\begin{equation}\label{eq:uvlf}
\Phi\equiv\frac{\mathrm{d}n}{\mathrm{d}M_{\mathrm{UV}}}=\frac{\mathrm{d}n}{\mathrm{d}M_{\mathrm{h}}}\left|\frac{\mathrm{d}M_{\mathrm{h}}}{\mathrm{d}M_*}\right|\left|\frac{\mathrm{d}M_*}{\mathrm{d}M_{\mathrm{UV}}}\right|,
\end{equation}
where $M_{\mathrm{h}}$ and $M_*$ denote the halo and stellar masses, respectively. The detailed calculation of the halo mass function, matter variance, stellar-to-halo mass relation, and conversion to the observed $M_{\mathrm{UV}}$ is given in Appendix~\ref{app:jwst}.}

{We construct the JWST likelihood directly in the $\log_{10}\Phi$ space. For each redshift bin, only detected data points with $\Phi_{\mathrm{obs}}>0$ are included in the $\chi^2$, while upper-limit points are not directly included in the $\chi^2$ evaluation. The likelihood is defined as
\begin{equation}
-2\ln\mathcal{L}_{\mathrm{JWST}}=\chi^2_{\mathrm{JWST}}=\sum_i\left[\frac{y_i^{\mathrm{obs}}-y_i^{\mathrm{th}}}{\sigma_i}\right]^2,
\end{equation}
where $y_i^{\mathrm{th}}=\log_{10}\Phi_i^{\mathrm{th}}$ and $y_i^{\mathrm{obs}}=\log_{10}\Phi_i^{\mathrm{obs}}$. To account for asymmetric observational uncertainties, the effective uncertainty in $\log_{10}\Phi$ is chosen according to the relative position of the theoretical prediction,
\begin{equation}
\sigma_i=\begin{cases}\sigma_i^{+},&y_i^{\mathrm{th}}\ge y_i^{\mathrm{obs}},\\\sigma_i^{-},&y_i^{\mathrm{th}}<y_i^{\mathrm{obs}},\end{cases}
\end{equation}
where the logarithmic uncertainties are obtained from the corresponding upper and lower observational errors. Thus, the likelihood is equivalent to using two half-Gaussian branches for the asymmetric errors. The six astrophysical parameters, $\{A,a,p,\log_{10}M_{\mathrm{c}},N,\beta\}$, are treated as free nuisance parameters and marginalized over in our likelihood analysis. The parameters $\{A,a,p\}$ characterize the halo mass function, while $\{\log_{10}M_{\mathrm{c}},N,\beta\}$ describe the stellar-to-halo mass relation. The detailed implementation is described in Appendix~\ref{app:jwst}.}

\begin{table*}[!htb]
\renewcommand\arraystretch{1.5}
\centering
\caption{{Cosmological parameter constraints and QDMAP $H_0$ tensions for $\Lambda$CDM and four EDE models. The QDMAP tension is calculated from the change in the best-fit $\chi^2$ between CMB+DESI+JWST and CMB+DESI+JWST+SH0ES.}}
\label{table1}
\resizebox{1.0\textwidth}{!}{
\setlength{\tabcolsep}{5pt}
\begin{tabular}{l|ccc|ccc|c}
\hline\hline
& \multicolumn{3}{c|}{CMB+DESI+JWST} & \multicolumn{3}{c|}{CMB+DESI+JWST+SH0ES} & \\
\cline{2-7}
Model & $H_0\,[\rm km\,s^{-1}\,Mpc^{-1}]$ & $f_{\mathrm{EDE}}(z_{\mathrm{c}})$ & $\log_{10}z_{\mathrm{c}}$ & $H_0\,[\rm km\,s^{-1}\,Mpc^{-1}]$ & $f_{\mathrm{EDE}}(z_{\mathrm{c}})$ & $\log_{10}z_{\mathrm{c}}$ & $Q_{\mathrm{DMAP}}$ \\
\hline
$\boldsymbol{\Lambda \mathrm{CDM}}$ & $68.21\pm0.26$ & $\text{---}$ & $\text{---}$ & $68.49\pm 0.25$ & $\text{---}$ & $\text{---}$ & $4.6\sigma$ \\
\textbf{Axion-EDE} & $71.58\pm1.05$ & $0.108^{+0.033}_{-0.025}$ & $3.57^{+0.04}_{-0.07}$ & $72.29\pm 0.72$ & $0.128^{+0.020}_{-0.018}$ & $3.58^{+0.03}_{-0.06}$ & $0.6\sigma$ \\
\textbf{$\alpha$-EDE} & $71.11^{+1.01}_{-1.25}$ & $0.086\pm0.031$ & $3.51^{+0.05}_{-0.03}$ & $72.18\pm 0.84$ & $0.113\pm 0.022$ & $3.52^{+0.04}_{-0.03}$ & $1.0\sigma$ \\
\textbf{RnR-EDE} & $70.53\pm0.98$ & $0.069\pm0.026$ & $3.49\pm0.06$ & $71.76^{+0.72}_{-0.80}$ & $0.100\pm 0.019$ & $3.49^{+0.04}_{-0.03}$ & $1.9\sigma$ \\
\textbf{AdS-EDE} & $70.71^{+0.56}_{-0.71}$ & $0.070^{+0.012}_{-0.016}$ & $3.42^{+0.01}_{-0.02}$ & $71.43\pm 0.62$ & $0.085\pm 0.014$ & $3.40\pm0.01$ & $1.8\sigma$ \\
\hline
\end{tabular}
}
\end{table*}

\subsection{Other Cosmological Datasets}

We consider several additional cosmological probes as follows.

\begin{itemize}

\item \textbf{CMB.} We employ the Planck 2018 CMB temperature and polarization power spectra~\citep{Efstathiou:2019mdh,Rosenberg:2022sdy}, together with high-multipole observations from ACT DR6~\citep{AtacamaCosmologyTelescope:2025blo} and SPT-3G~\citep{SPT-3G:2025bzu}. We impose strict multipole cutoffs on the Planck data ($\ell<1000$ for TT and $\ell<600$ for TE and EE) to prevent mode double-counting~\citep{AtacamaCosmologyTelescope:2025blo}. We further include the joint CMB lensing potential power spectrum reconstructed from Planck PR4, ACT DR6, and SPT-3G~\citep{Carron:2022eyg,ACT:2023kun,ACT:2023dou,ACT:2025qjh}.

\item \textbf{DESI.} We utilize the BAO measurements provided by DESI DR2~\citep{DESI:2025zgx}, including the transverse comoving distance $D_{\mathrm{M}}/r_{\mathrm{d}}$, the angle-averaged distance $D_{\mathrm{V}}/r_{\mathrm{d}}$, and the Hubble distance $D_{\mathrm{H}}/r_{\mathrm{d}}$.

\item \textbf{SH0ES.} We adopt the local distance-ladder measurement $H_0=73.04\pm1.04~\mathrm{km\,s^{-1}\,Mpc^{-1}}$ from SH0ES~\citep{Riess:2021jrx} as the local $H_0$ prior.

\end{itemize}

\subsection{Bayesian Inference}

We perform the cosmological parameter estimation using the public package \texttt{Cobaya}~\citep{Torrado:2020dgo}. {The theoretical observables are computed using the modified Einstein-Boltzmann solver \texttt{CAMB}~\citep{Lewis:1999bs,Howlett:2012mh}, and the resulting chains are post-processed with \texttt{GetDist}~\citep{Lewis:2019xzd}. Specifically, the four EDE models are implemented by modifying the \texttt{DarkEnergyQuintessence} module in \texttt{CAMB} to solve the corresponding scalar-field dynamics and to consistently compute the associated linear perturbations.} The relative fitting performance and model selection are evaluated using the minimum chi-squared difference $\Delta\chi^2$ relative to $\Lambda$CDM and the Deviance Information Criterion (DIC) defined as $\mathrm{DIC}\equiv\bar{\chi}^2+p_D$, where $\bar{\chi}^2$ represents the mean of the posterior chi-squared distribution and $p_D\equiv\bar{\chi}^2-\chi^2(\bar{\boldsymbol{\theta}})$ denotes the effective number of parameters evaluated at the mean parameter vector $\bar{\boldsymbol{\theta}}$~\citep{Spiegelhalter:2002yvw}.

The baseline parameter space is $\boldsymbol{\theta}_{\Lambda\mathrm{CDM}}=\{\Omega_{\mathrm{b}}h^2,\Omega_{\mathrm{c}}h^2,100\theta_{\mathrm{MC}},\tau,\ln(10^{10}A_{\mathrm{s}}),n_{\mathrm{s}}\}$. The extension of this parameter space varies depending on the specific EDE physics. The Axion-EDE, $\alpha$-EDE, and RnR-EDE frameworks expand the baseline by adding three sampling parameters $\boldsymbol{\theta}_{\mathrm{EDE}}=\{\boldsymbol{\theta}_{\Lambda\mathrm{CDM}},f_{\mathrm{EDE}},\log_{10}z_{\mathrm{c}},\phi_{\mathrm{i}}\}$. Conversely, the AdS-EDE framework introduces only two additional sampling parameters $\boldsymbol{\theta}_{\mathrm{AdS}}=\{\boldsymbol{\theta}_{\Lambda\mathrm{CDM}},f_{\mathrm{EDE}},\log_{10}z_{\mathrm{c}}\}$. Following the methodology of \citet{Ye:2020btb}, we fix $\alpha_{\mathrm{AdS}}=3.79\times10^{-4}$ to avoid numerical integration divergence in the background solvers and ensure stable chain convergence.

\begin{figure*}[htb]
\centering
\includegraphics[width=1.0\textwidth]{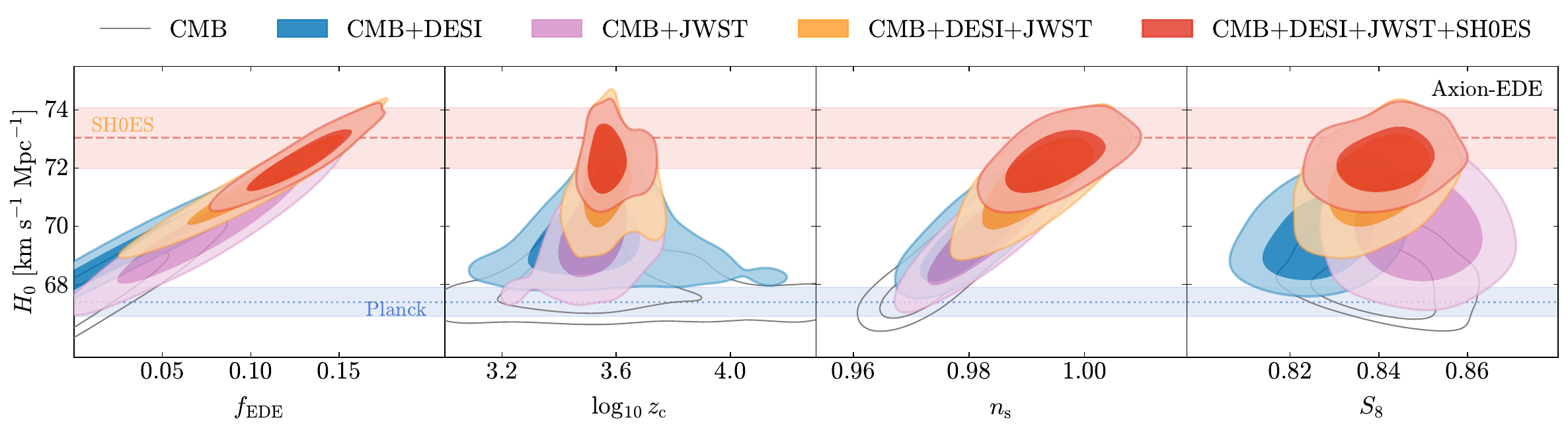}
\caption{
The $1\sigma$ and $2\sigma$ marginalized credible-interval contours for $H_0$ with $f_{\mathrm{EDE}}$, $\log_{10} z_\mathrm{c}$, $n_\mathrm{s}$, and $S_8$ in the Axion-EDE model from the CMB, DESI, JWST, and SH0ES data. The horizontal shaded bands represent the local $H_0$ measurement from SH0ES (light red) and the CMB-inferred value from Planck (light blue).
}
\label{fig1}
\end{figure*}

\section{Results and discussion}\label{sec3}

{As shown in Table~\ref{table1}, within the $\Lambda$CDM model, the CMB+DESI+JWST data favor a lower $H_0$ of $68.21\pm0.26~\mathrm{km\,s^{-1}\,Mpc^{-1}}$. To avoid the circularity associated with directly comparing a posterior that already includes SH0ES with the SH0ES measurement, we adopt the $Q_{\rm DMAP}$ metric~\citep{Schoneberg:2021qvd,Bella:2026zuk}. Specifically, we define $\Delta\chi^2_{H_0}\equiv\chi^2_{\min}(D+\mathrm{SH0ES})-\chi^2_{\min}(D)$, where $D$ denotes the cosmological dataset without SH0ES, and quantify the residual $H_0$ tension as $Q_{\rm DMAP}\equiv\sqrt{\Delta\chi^2_{H_0}}$. The corresponding CMB+DESI+JWST+SH0ES constraints are also shown in Table~\ref{table1}. Compared with the SH0ES-free results, the addition of the SH0ES prior shifts $H_0$ toward the local measurement and generally leads to a moderately larger EDE fraction, while the preferred $\log_{10}z_{\mathrm{c}}$ remains nearly unchanged. For the CMB+DESI+JWST data, the Axion-EDE model yields $H_0=71.58\pm1.05~\mathrm{km\,s^{-1}\,Mpc^{-1}}$, with $f_{\mathrm{EDE}}(z_{\mathrm{c}})=0.108^{+0.033}_{-0.025}$ and $\log_{10}z_{\mathrm{c}}=3.57^{+0.04}_{-0.07}$, reducing the residual $H_0$ tension to $0.6\sigma$. The $\alpha$-EDE, RnR-EDE, and AdS-EDE models similarly reduce the $H_0$ tension to $1.0\sigma$, $1.9\sigma$, and $1.8\sigma$, respectively. These results demonstrate that the high-redshift JWST data play an important role in alleviating the $H_0$ tension across different EDE realizations. The detailed parameter constraints and QDMAP results for the other datasets are presented in Appendix~\ref{app:results}.}

In Figure~\ref{fig1}, we present the two-dimensional posterior contour plots of $H_0$ with $f_{\mathrm{EDE}}$, $\log_{10} z_{\mathrm{c}}$, $n_{\mathrm{s}}$, and $S_8$ in the Axion-EDE model. A significant positive correlation between $f_{\mathrm{EDE}}$ and $H_0$ is clearly observable. This indicates that the peak fractional density of the EDE is directly associated with the magnitude of the $H_0$ enhancement. Simultaneously, the scalar spectral index $n_{\mathrm{s}}$ exhibits a distinct positive correlation with $H_0$, which is a general conclusion within EDE scenarios~\citep{Knox:2019rjx,Niedermann:2020dwg,Ye:2021nej,Jiang:2022uyg,Jiang:2024nha}. In particular, fully reconciling $H_0$ with the local SH0ES measurement requires a spectral index much closer to the Harrison-Zel'dovich spectrum ($n_{\mathrm{s}}\simeq1$), which would have profound implications for inflationary theories~\citep{Ye:2021nej,Jiang:2022uyg}. Furthermore, the constraints on the structure growth amplitude $S_8$ reveal obvious discrepancies among different datasets. Compared to CMB alone, the inclusion of JWST data tends to yield a higher $S_8$ value, while the DESI data favor a lower $S_8$, which relates to the recently debated $\Omega_{\mathrm{m}}$ discrepancy between CMB and DESI~\citep{Colgain:2024mtg,Tang:2024lmo,Colgain:2024xqj,Wang:2025znm}. Importantly, the $S_8$ value derived from the combined CMB+DESI+JWST dataset remains largely consistent with the constraints from CMB alone. This demonstrates that while the EDE models effectively alleviate the $H_0$ tension, they do not exacerbate the $S_8$ tension. \footnote{It is worth noting that the $S_8$ tension between KiDS-Legacy cosmic shear data and Planck is only $0.73\sigma$~\citep{Wright:2025xka}, whereas the recently released DES Y6 observations yield $S_8=0.789\pm0.012$, remaining systematically lower than the Planck prediction and presenting a tension of approximately $2.6\sigma$~\citep{DES:2026fyc}. This discrepancy in measurements suggests that systematic errors specific to weak lensing observations, including photometric redshift calibration, intrinsic alignments, and complex baryonic feedback, might be potential sources of the current $S_8$ tension.}

\begin{figure}[htbp]
\centering
\includegraphics[width=0.49\textwidth]{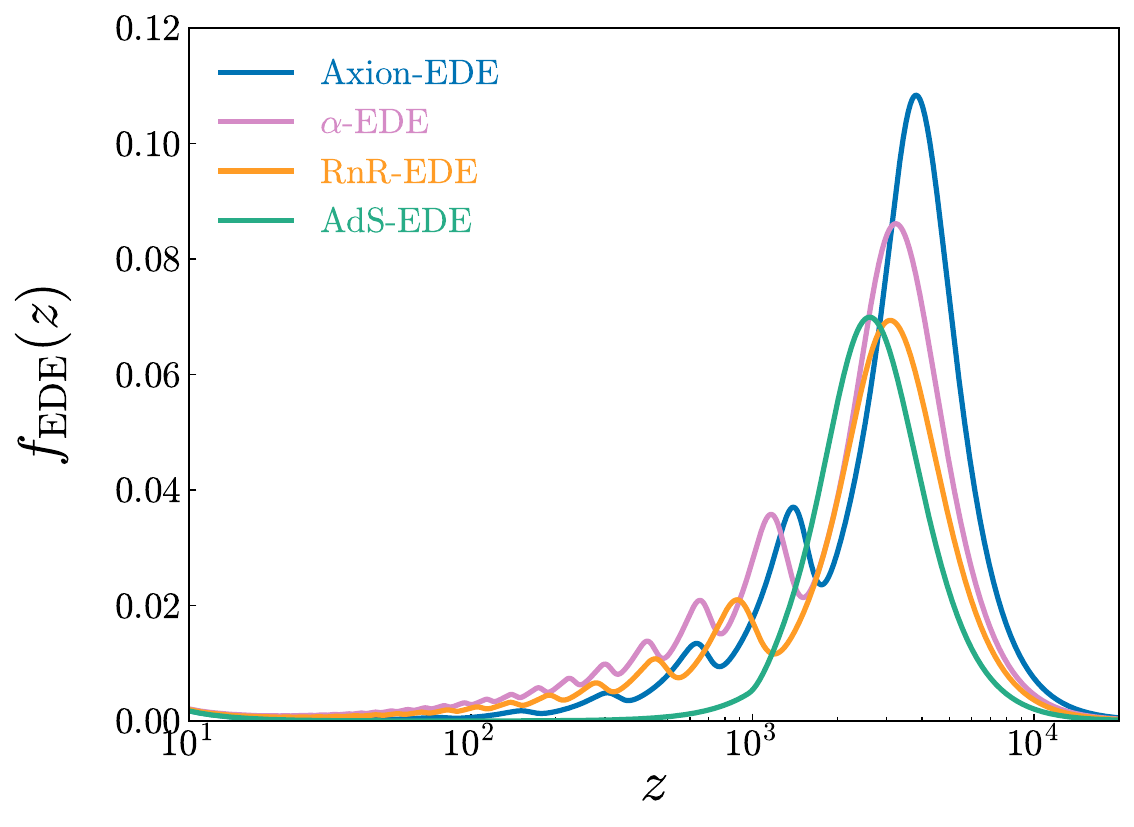}
\caption{
The evolution of the fractional energy density $f_{\mathrm{EDE}}(z)$ with respect to redshift $z$ for the four EDE models using the best-fit results obtained from CMB+DESI+JWST.
}
\label{fig2}
\end{figure}

\begin{figure*}[htbp]
\centering
\includegraphics[width=1.0\textwidth]{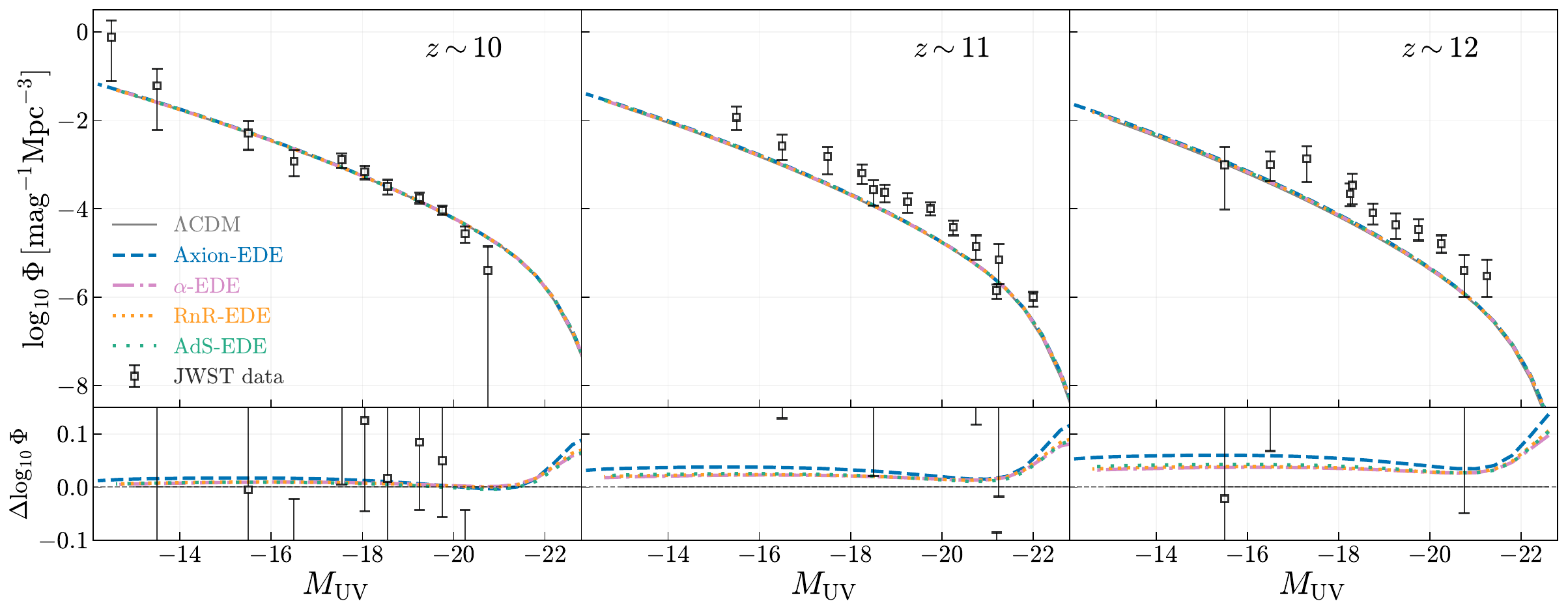}
\caption{
The theoretical predictions of the ultraviolet luminosity functions $\log_{10}\Phi$ at high redshifts ($z \sim 10$, $11$, and $12$) in four EDE models (upper panels), along with their residuals $\Delta\log_{10}\Phi$ relative to the best-fit $\Lambda$CDM model (lower panels). The data points with error bars represent the high-redshift observations from JWST.
}
\label{fig3}
\end{figure*}

In Figure~\ref{fig2}, we display the evolution behavior of $f_{\mathrm{EDE}}(z)$ for the four EDE models under their respective best-fit parameters. It can be observed that these models follow the same physical picture in terms of their overall dynamics. The scalar field energy rapidly increases to a peak near the critical redshift $z_{\mathrm{c}}$, thereby effectively reducing the early sound horizon, and is subsequently diluted at a rapid pace. Specifically, the Axion-EDE model exhibits the highest peak fraction and is followed by a noticeable high-frequency oscillatory decay. The peak values for the $\alpha$-EDE and RnR-EDE models are slightly lower with a relatively broader redshift distribution of energy injection, and they display oscillatory decay tails with distinct characteristics. It is noteworthy that the AdS-EDE model demonstrates a completely different decay behavior compared to the other three models during the evolution phase at redshifts $z<1500$. After passing the peak, the scalar field in the AdS-EDE model rolls into the Anti-de Sitter potential well with a negative depth. During this phase, the scalar field is in a kinetic energy dominated state, causing the equation of state parameter to sharply increase to $w>1$. Consequently, its energy density $\rho_\phi$ redshifts and decays at an extremely high rate ($\propto a^{-3(1+w)}$) as the universe expands. This energy dissipation mechanism ensures that while providing sufficient early energy injection to alleviate the $H_0$ tension, the model can maximally avoid unnecessary interference with the post-recombination cosmic evolution such as the formation of large-scale structures.

In Figure~\ref{fig3}, the theoretical predictions of the ultraviolet luminosity functions at high redshifts ($z\sim10$, $11$, and $12$) under different EDE models, along with their residuals relative to the $\Lambda$CDM model, are presented. It can be clearly observed that at $z\sim10$, the theoretical curves are in basic agreement with the JWST observational data, whereas at $z\sim11$ and $z\sim12$, the actual JWST data points are significantly higher than the theoretical predictions. Simultaneously, the modification effect of EDE on the theoretical curves becomes more prominent with increasing redshift. This phenomenon is primarily attributed to the relatively higher residual density fraction of EDE at higher redshifts, which subsequently exerts a stronger influence on the formation of early large-scale structures and the matter clustering rate. Specifically, within the redshift range at $z\sim12$, the Axion-EDE model enhances the theoretical predictions by approximately $5\%$ at the bright end of the ultraviolet luminosity function and up to about $10\%$ at the faint end compared to the standard $\Lambda$CDM model. This indicates that the introduction of EDE models helps improve the fit to the high-redshift massive galaxy observations from JWST to some extent. By contrast, the improvements offered by the other three EDE models remain relatively limited.

\begin{table}[htbp]
\renewcommand\arraystretch{1.4}
\centering
\caption{The individual $\Delta\chi^2$ (CMB, DESI, and JWST), total $\Delta\chi^2$, and $\Delta\mathrm{DIC}$ for the four EDE models relative to $\Lambda$CDM, utilizing the CMB+DESI+JWST data. Negative values indicate a better fit compared to $\Lambda$CDM.}
\label{table2}
\resizebox{0.49\textwidth}{!}{
\setlength{\tabcolsep}{4pt}
\begin{tabular}{lccccc}
\hline\hline
Model & $\Delta\chi^2_{\rm CMB}$ & $\Delta\chi^2_{\rm DESI}$ & $\Delta\chi^2_{\rm JWST}$ & $\Delta \chi^2_{\rm tot}$ & $\Delta\mathrm{DIC}$ \\
\hline
Axion-EDE   & $ -3.72$ & $ -1.05$ & $-13.49$ & $-18.26$ & $-11.89$ \\
$\alpha$-EDE & $ -2.10$ & $ -1.34$ & $ -5.37$ & $ -8.82$ & $ -7.93$ \\
RnR-EDE     & $ -1.07$ & $ -0.99$ & $ -6.35$ & $ -8.41$ & $ -5.65$ \\
AdS-EDE     & $ -0.99$ & $ -1.80$ & $ -7.52$ & $-10.31$ & $ -9.06$ \\
\hline
\end{tabular}
}
\end{table}

To quantitatively compare the goodness-of-fit among the models, we report the individual $\Delta\chi^2$ components, $\Delta\chi^2_{\mathrm{tot}}$, and $\Delta\mathrm{DIC}$ for each EDE model relative to the $\Lambda$CDM model using CMB+DESI+JWST in Table~\ref{table2}. All four EDE models exhibit a better goodness-of-fit than $\Lambda$CDM, and this fitting improvement is reflected across every individual observational data, with the enhancement for the JWST data being the most significant. This strongly demonstrates that the CMB+DESI+JWST data possesses a clear statistical preference for the EDE scenarios. It is worth noting that the introduction of the EDE component does not impose any adverse effects on the fit to the CMB data, and its corresponding chi-squared component is actually reduced. This result indicates that while the $H_0$ tension is substantially alleviated, the EDE models do not compromise the high-precision fit to the CMB observations. Specifically, the Axion-EDE model, which yields the most significant mitigation effect on the $H_0$ tension, obtains the largest improvement in the goodness-of-fit, resulting in $\Delta\chi^2_{\mathrm{tot}}=-18.26$ and $\Delta\mathrm{DIC}=-11.89$. The AdS-EDE model closely follows in overall fitting performance, while the statistical improvements brought by the $\alpha$-EDE and RnR-EDE models are relatively limited.

\section{Conclusion}\label{sec4}

This letter aims to investigate how the independent observational signals of early universe structure formation provided by JWST influence the evolutionary dynamics of EDE and effectively alleviate the $H_0$ tension. The results indicate that the high-redshift JWST data play a crucial role in alleviating the $H_0$ tension. In the Axion-EDE model, the CMB+DESI+JWST dataset significantly raises $H_0$ to $71.58\pm1.05\ \mathrm{km\,s^{-1}\,Mpc^{-1}}$, reducing the $H_0$ tension to $Q_{\mathrm{DMAP}}=0.6\sigma$. The corresponding $H_0$ tensions for the $\alpha$-EDE, RnR-EDE, and AdS-EDE models are reduced to $1.0\sigma$, $1.9\sigma$, and $1.8\sigma$, respectively. These results demonstrate that the inclusion of JWST data can substantially alleviate the $H_0$ tension across different EDE realizations. Regarding the fit to the JWST data, the theoretical curve of the Axion-EDE model at $z \sim 12$ elevates the predicted values at the bright and faint ends of the ultraviolet luminosity function by approximately $5\%$ and $10\%$, which helps improve the agreement with JWST. Furthermore, we quantitatively compare the goodness of fit and find that the Axion-EDE model exhibits a statistical performance superior to the standard cosmological model, achieving significant improvements of $\Delta\chi^2_{\mathrm{tot}} = -18.26$ and $\Delta\mathrm{DIC} = -11.89$.

In summary, this research emphasizes the core value of combining high-redshift galaxy observations with traditional CMB and large-scale structure data in testing early universe cosmological models. In the future, with the JWST acquiring more deep field observational data, along with the successful advancement of next-generation CMB experiments and large-scale galaxy surveys, we will be able to test these EDE physical scenarios with unprecedented precision and potentially provide a further answer to whether the $H_0$ tension originates from systematic errors or new physics.

\section*{Acknowledgments}
We thank Jun-Qian Jiang, Yun-He Li and Yi-Min Zhang for their helpful discussions. This work was supported by the National Natural Science Foundation of China (Grants Nos. 12533001, 12575049, and 12473001), the National SKA Program of China (Grants Nos. 2022SKA0110200 and 2022SKA0110203), the China Manned Space Program (Grant No. CMS-CSST-2025-A02), the 111 Project (Grant No. B16009), and the Natural Science Foundation of Shanghai (Grant No. 24ZR1424600). This work is based [in part] on observations made with the NASA/ESA/CSA James Webb Space Telescope (JWST) and Hubble Space Telescope (HST). The data were obtained from the Mikulski Archive for Space Telescopes (MAST) at the Space Telescope Science Institute, which is operated by the Association of Universities for Research in Astronomy, Inc., under NASA contract NAS 5-03127 for JWST and NAS 5-26555 for HST. These observations are associated with programs 3293, 1837, 1180, 1210, and 2079, and accessed via \dataset[10.17909/je9x-d314]{https://doi.org/10.17909/je9x-d314} and \dataset[10.17909/gbkc-2b17]{https://doi.org/10.17909/gbkc-2b17}. We also acknowledge data obtained from the ESO Science Archive Facility with \dataset[10.18727/archive/52]{https://doi.org/10.18727/archive/52}.

\bibliography{main}

\appendix

\section{{Detailed Modeling of the JWST UV Luminosity Function}}
\label{app:jwst}

{In this appendix, we provide the detailed calculation of the theoretical ultraviolet luminosity function (UVLF). The first term $\mathrm{d}n/\mathrm{d}M_{\mathrm{h}}$ in Equation~\eqref{eq:uvlf} is the halo mass function calculated via the Sheth-Mo-Tormen formalism~\citep{Sheth:1999su},
\begin{equation}
f_{\mathrm{SMT}}(\sigma)=A\sqrt{\frac{2a}{\pi}}\nu\exp\left(-\frac{a\nu^2}{2}\right)\left[1+(a\nu^2)^{-p}\right],
\end{equation}
where $\nu\equiv\delta_{\mathrm{crit}}/\sigma$ with the critical collapse threshold $\delta_{\mathrm{crit}}=1.686$, and $\{A,a,p\}$ are free parameters. The corresponding halo mass function is obtained from
\begin{equation}
\frac{\mathrm{d}n}{\mathrm{d}M_{\mathrm{h}}}=\frac{\rho_{\mathrm{m}}}{M_{\mathrm{h}}}f_{\mathrm{SMT}}(\sigma)\left|\frac{\mathrm{d}\ln\sigma^{-1}}{\mathrm{d}M_{\mathrm{h}}}\right|,
\end{equation}
where $\sigma$ represents the matter density variance evaluated at the comoving scale $R$ corresponding to the halo mass,
\begin{equation}
\sigma^2(R)=\int_0^\infty\frac{k^2\mathrm{d}k}{2\pi^2}P(k)|W(kR)|^2,
\end{equation}
where $P(k)$ is the matter power spectrum and $W(x)=3(\sin x-x\cos x)/x^3$ is the spherical top-hat window function.}

{The stellar mass is linked to the halo mass through a double power-law stellar-to-halo mass relation parameterized as~\citep{Stefanon:2021},
\begin{equation}
\frac{M_*}{M_{\mathrm{h}}}=2N\left[\left(\frac{M_{\mathrm{h}}}{M_{\mathrm{c}}}\right)^{-\beta}+\left(\frac{M_{\mathrm{h}}}{M_{\mathrm{c}}}\right)^\gamma\right]^{-1},
\end{equation}
where $N$ defines the normalization amplitude, $M_{\mathrm{c}}$ is the characteristic halo scale, and $\beta$ describes the low-mass slope. The high-mass slope is fixed to $\gamma=-0.01$ following \citet{CEERSTeam:2023qgy,Scholtz:2023}.}

{The conversion from stellar mass to the observed ultraviolet magnitude includes the dust attenuation correction calibrated by the ultraviolet continuum slope. The intrinsic ultraviolet magnitude is related to the stellar mass through
\begin{equation}
M_{\mathrm{UV}}^{\mathrm{int}}=-20.5+\frac{\log_{10}(M_*)-a_0}{a_1},
\end{equation}
where $a_0$ and $a_1$ are redshift-dependent parameters given in \citet{Stefanon:2021}. The observed ultraviolet magnitude is obtained after applying the attenuation correction,
\begin{equation}
M_{\mathrm{UV}}=M_{\mathrm{UV}}^{\mathrm{int}}+A_{1600},
\end{equation}
where the dust attenuation is calibrated through the ultraviolet continuum slope $\beta_{\mathrm{slope}}$ via the empirical relations~\citep{Meurer:1999jj,Cullen:2022},
\begin{equation}
A_{1600}=4.43+1.99\beta_{\mathrm{slope}},
\end{equation}
and
\begin{equation}
\beta_{\mathrm{slope}}=-0.17\frac{\log_{10}(M_*)-a_0}{a_1}-5.40.
\end{equation}}

\section{{Full Cosmological Constraints and Statistical Results}}
\label{app:results}

{In this appendix, we provide the full cosmological parameter constraints and statistical results for all considered datasets and models. The constraints without and with the SH0ES calibration are summarized in Tables~\ref{table_ede_params_noSH0ES} and \ref{table_ede_params_SH0ES}, respectively. The corresponding quantification of the $H_0$ tension using the $\Delta\chi^2_{H_0}$ and $Q_{\mathrm{DMAP}}$ statistics is presented in Table~\ref{table_qdmap}.}

{For comparison with the impact of JWST observations, we include two supernova datasets with different distance calibrations:}
\begin{itemize}
\item {\textbf{PantheonPlus (PP).} We use the PantheonPlus compilation containing 1550 spectroscopically confirmed Type Ia supernovae over the redshift range $0.01<z<2.26$~\citep{Brout:2022vxf}. We label this dataset as ``\textbf{PP}''.}

\item {\textbf{PantheonPlusSH0ES (PPS).} We further consider the PantheonPlusSH0ES compilation, which combines the PantheonPlus sample with the SH0ES Cepheid calibration and provides an absolute distance anchor from the local distance ladder~\citep{Brout:2022vxf,Riess:2021jrx}. We label this dataset as ``\textbf{PPS}''.}
\end{itemize}

\begin{table*}[!htb]
\renewcommand\arraystretch{1.3}
\centering
\caption{{Marginalized cosmological parameter constraints for $\Lambda$CDM and four EDE models from CMB, DESI, JWST, and PP data without the SH0ES prior.}}
\label{table_ede_params_noSH0ES}
\resizebox{\textwidth}{!}{
\setlength{\tabcolsep}{5pt}
\begin{tabular}{lccccccc}
\hline\hline
Model / Data & $H_0\,[\rm km\,s^{-1}\,Mpc^{-1}]$ & $A_s\,(\times 10^{-9})$ & $n_s$ & $\Omega_m$ & $S_8$ & $f_{\mathrm{EDE}}(z_c)$ & $\log_{10}z_{\mathrm{c}}$ \\
\hline
\multicolumn{8}{l}{$\boldsymbol{\Lambda\mathrm{CDM}}$} \\
CMB & $67.24\pm0.39$ & $2.111\pm0.027$ & $0.9691^{+0.0035}_{-0.0032}$ & $0.3168\pm0.0056$ & $0.8373\pm0.0089$ & $\text{---}$ & $\text{---}$ \\
CMB+DESI & $68.20\pm0.26$ & $2.149^{+0.025}_{-0.029}$ & $0.9740\pm0.0029$ & $0.3031^{+0.0033}_{-0.0037}$ & $0.8199\pm0.0067$ & $\text{---}$ & $\text{---}$ \\
CMB+JWST & $67.23\pm0.41$ & $2.130^{+0.025}_{-0.029}$ & $0.9705\pm0.0034$ & $0.3170\pm0.0059$ & $0.8420\pm0.0088$ & $\text{---}$ & $\text{---}$ \\
CMB+DESI+JWST & $68.21\pm0.26$ & $2.175\pm0.029$ & $0.9760\pm0.0030$ & $0.3030\pm0.0036$ & $0.8253\pm0.0066$ & $\text{---}$ & $\text{---}$ \\
CMB+DESI+PP & $68.10\pm0.26$ & $2.143^{+0.025}_{-0.028}$ & $0.9736\pm0.0030$ & $0.3046\pm0.0035$ & $0.8216\pm0.0067$ & $\text{---}$ & $\text{---}$ \\
\hline
\multicolumn{8}{l}{$\boldsymbol{\mathrm{Axion\mathchar`-EDE}}$} \\
CMB & $68.15^{+0.61}_{-1.02}$ & $2.116\pm0.027$ & $0.9731^{+0.0047}_{-0.0059}$ & $0.3146\pm0.0062$ & $0.8403\pm0.0092$ & $<0.069\,(2\sigma)$ & $3.55^{+0.13}_{-0.25}$ \\
CMB+DESI & $69.66^{+0.78}_{-1.17}$ & $2.150\pm0.026$ & $0.9807^{+0.0049}_{-0.0064}$ & $0.3013\pm0.0039$ & $0.8270\pm0.0081$ & $0.048^{+0.013}_{-0.048}$ & $3.54^{+0.10}_{-0.16}$ \\
CMB+JWST & $69.95^{+1.11}_{-1.34}$ & $2.148^{+0.027}_{-0.031}$ & $0.9839\pm0.0070$ & $0.3093\pm0.0067$ & $0.8486\pm0.0091$ & $0.078\pm0.033$ & $3.53\pm0.09$ \\
CMB+DESI+JWST & $71.58\pm1.05$ & $2.180^{+0.025}_{-0.029}$ & $0.9918^{+0.0059}_{-0.0066}$ & $0.2981\pm0.0038$ & $0.8397^{+0.0081}_{-0.0069}$ & $0.108^{+0.033}_{-0.025}$ & $3.57^{+0.04}_{-0.07}$ \\
CMB+DESI+PP & $69.53^{+0.74}_{-1.23}$ & $2.148^{+0.024}_{-0.027}$ & $0.9799^{+0.0050}_{-0.0064}$ & $0.3025\pm0.0036$ & $0.8284\pm0.0082$ & $<0.104\,(2\sigma)$ & $3.53^{+0.10}_{-0.17}$ \\
\hline
\multicolumn{8}{l}{$\boldsymbol{\mathrm{\alpha\mathchar`-\rm EDE}}$} \\
CMB & $68.34^{+0.66}_{-1.12}$ & $2.122^{+0.027}_{-0.031}$ & $0.9724\pm0.0045$ & $0.3146\pm0.0062$ & $0.8395\pm0.0093$ & $<0.076\,(2\sigma)$ & $3.49^{+0.12}_{-0.10}$ \\
CMB+DESI & $69.85^{+0.87}_{-1.28}$ & $2.162\pm0.028$ & $0.9787\pm0.0042$ & $0.3010\pm0.0038$ & $0.8246\pm0.0072$ & $0.050^{+0.023}_{-0.040}$ & $3.49^{+0.08}_{-0.06}$ \\
CMB+JWST & $69.55\pm1.11$ & $2.157\pm0.030$ & $0.9783\pm0.0049$ & $0.3117\pm0.0063$ & $0.8462\pm0.0094$ & $0.065^{+0.031}_{-0.027}$ & $3.50\pm0.06$ \\
CMB+DESI+JWST & $71.11^{+1.01}_{-1.25}$ & $2.199^{+0.029}_{-0.033}$ & $0.9842^{+0.0042}_{-0.0050}$ & $0.2993\pm0.0038$ & $0.8339\pm0.0073$ & $0.086\pm0.031$ & $3.51^{+0.05}_{-0.03}$ \\
CMB+DESI+PP & $69.57^{+0.82}_{-1.18}$ & $2.156\pm0.029$ & $0.9778\pm0.0043$ & $0.3025\pm0.0037$ & $0.8258\pm0.0075$ & $<0.093\,(2\sigma)$ & $3.48^{+0.10}_{-0.06}$ \\
\hline
\multicolumn{8}{l}{$\boldsymbol{\mathrm{RnR\mathchar`-EDE}}$} \\
CMB & $68.15^{+0.62}_{-0.96}$ & $2.120\pm0.027$ & $0.9727^{+0.0047}_{-0.0054}$ & $0.3154\pm0.0057$ & $0.8408\pm0.0092$ & $<0.066\,(2\sigma)$ & $3.46^{+0.15}_{-0.11}$ \\
CMB+DESI & $69.25^{+0.54}_{-0.98}$ & $2.156^{+0.026}_{-0.029}$ & $0.9784^{+0.0043}_{-0.0054}$ & $0.3022\pm0.0036$ & $0.8245^{+0.0072}_{-0.0080}$ & $<0.075\,(2\sigma)$ & $3.48^{+0.13}_{-0.12}$ \\
CMB+JWST & $69.48^{+0.99}_{-1.14}$ & $2.154^{+0.027}_{-0.031}$ & $0.9810\pm0.0058$ & $0.3134\pm0.0061$ & $0.8510\pm0.0098$ & $0.065\pm0.027$ & $3.48\pm0.06$ \\
CMB+DESI+JWST & $70.53\pm0.98$ & $2.191\pm0.029$ & $0.9866\pm0.0054$ & $0.3009\pm0.0036$ & $0.8360\pm0.0080$ & $0.069\pm0.026$ & $3.49\pm0.06$ \\
CMB+DESI+PP & $69.09^{+0.56}_{-0.90}$ & $2.153\pm0.028$ & $0.9777^{+0.0043}_{-0.0049}$ & $0.3033\pm0.0036$ & $0.8256\pm0.0074$ & $<0.069\,(2\sigma)$ & $3.48\pm0.16$ \\
\hline
\multicolumn{8}{l}{$\boldsymbol{\mathrm{AdS\mathchar`-EDE}}$} \\
CMB & $69.39^{+0.53}_{-0.61}$ & $2.121\pm0.027$ & $0.9800\pm0.0037$ & $0.3107\pm0.0059$ & $0.8461\pm0.0092$ & $0.055^{+0.005}_{-0.011}$ & $3.43\pm0.01$ \\
CMB+DESI & $70.30^{+0.43}_{-0.59}$ & $2.145\pm0.025$ & $0.9838\pm0.0035$ & $0.3004\pm0.0036$ & $0.8330\pm0.0070$ & $0.059^{+0.008}_{-0.014}$ & $3.42^{+0.01}_{-0.02}$ \\
CMB+JWST & $69.73^{+0.63}_{-0.79}$ & $2.139\pm0.027$ & $0.9824\pm0.0039$ & $0.3098\pm0.0059$ & $0.8510\pm0.0088$ & $0.062^{+0.009}_{-0.016}$ & $3.42^{+0.01}_{-0.02}$ \\
CMB+DESI+JWST & $70.71^{+0.56}_{-0.71}$ & $2.165\pm0.027$ & $0.9866\pm0.0035$ & $0.3000\pm0.0036$ & $0.8406\pm0.0072$ & $0.070^{+0.012}_{-0.016}$ & $3.42^{+0.01}_{-0.02}$ \\
CMB+DESI+PP & $70.23^{+0.42}_{-0.63}$ & $2.141\pm0.025$ & $0.9834\pm0.0035$ & $0.3016\pm0.0035$ & $0.8347\pm0.0071$ & $0.060^{+0.007}_{-0.014}$ & $3.42^{+0.01}_{-0.02}$ \\
\hline
\end{tabular}
}
\end{table*}

\begin{table*}[!htb]
\renewcommand\arraystretch{1.3}
\centering
\caption{{Marginalized cosmological parameter constraints for $\Lambda$CDM and the four EDE models from the CMB, DESI, JWST, and PPS data including the SH0ES prior.}}
\label{table_ede_params_SH0ES}
\resizebox{\textwidth}{!}{
\setlength{\tabcolsep}{5pt}
\begin{tabular}{lccccccc}
\hline\hline
Model / Data & $H_0\,[\rm km\,s^{-1}\,Mpc^{-1}]$ & $A_s\,(\times 10^{-9})$ & $n_s$ & $\Omega_m$ & $S_8$ & $f_{\mathrm{EDE}}(z_c)$ & $\log_{10}z_{\mathrm{c}}$ \\
\hline
\multicolumn{8}{l}{$\boldsymbol{\Lambda\mathrm{CDM}}$} \\
CMB+SH0ES & $68.00\pm0.38$ & $2.138\pm0.026$ & $0.9726^{+0.0031}_{-0.0026}$ & $0.3062\pm0.0052$ & $0.8234^{+0.0079}_{-0.0088}$ & $\text{---}$ & $\text{---}$ \\
CMB+DESI+SH0ES & $68.48\pm0.24$ & $2.160\pm0.027$ & $0.9751\pm0.0028$ & $0.2996\pm0.0032$ & $0.8152\pm0.0066$ & $\text{---}$ & $\text{---}$ \\
CMB+JWST+SH0ES & $68.04^{+0.68}_{-0.44}$ & $2.168^{+0.032}_{-0.039}$ & $0.9750^{+0.0032}_{-0.0043}$ & $0.3055^{+0.0066}_{-0.0051}$ & $0.8285^{+0.0070}_{-0.0085}$ & $\text{---}$ & $\text{---}$ \\
CMB+DESI+JWST+SH0ES & $68.49\pm0.25$ & $2.190\pm0.030$ & $0.9774\pm0.0030$ & $0.2992\pm0.0034$ & $0.8209\pm0.0065$ & $\text{---}$ & $\text{---}$ \\
CMB+DESI+PPS & $68.47\pm0.26$ & $2.157^{+0.027}_{-0.032}$ & $0.9753^{+0.0030}_{-0.0027}$ & $0.2997\pm0.0034$ & $0.8151\pm0.0067$ & $\text{---}$ & $\text{---}$ \\
\hline
\multicolumn{8}{l}{$\boldsymbol{\mathrm{Axion\mathchar`-EDE}}$} \\
CMB+SH0ES & $70.33^{+0.66}_{-0.46}$ & $2.135\pm0.026$ & $0.9830\pm0.0046$ & $0.3037\pm0.0055$ & $0.8347\pm0.0094$ & $0.079^{+0.020}_{-0.016}$ & $3.54^{+0.05}_{-0.08}$ \\
CMB+DESI+SH0ES & $71.55\pm0.84$ & $2.160^{+0.024}_{-0.022}$ & $0.9899^{+0.0061}_{-0.0076}$ & $0.2969^{+0.0037}_{-0.0032}$ & $0.8320\pm0.0084$ & $0.104\pm0.025$ & $3.56^{+0.07}_{-0.08}$ \\
CMB+JWST+SH0ES & $71.86^{+0.85}_{-0.71}$ & $2.169^{+0.032}_{-0.029}$ & $0.9931\pm0.0053$ & $0.3010\pm0.0053$ & $0.8466\pm0.0085$ & $0.124^{+0.021}_{-0.016}$ & $3.57\pm0.05$ \\
CMB+DESI+JWST+SH0ES & $72.29\pm0.72$ & $2.183^{+0.025}_{-0.028}$ & $0.9949^{+0.0051}_{-0.0059}$ & $0.2965\pm0.0035$ & $0.8416^{+0.0076}_{-0.0066}$ & $0.128^{+0.020}_{-0.018}$ & $3.58^{+0.03}_{-0.06}$ \\
CMB+DESI+PPS & $71.93\pm0.73$ & $2.161\pm0.025$ & $0.9905^{+0.0047}_{-0.0055}$ & $0.2977\pm0.0032$ & $0.8367\pm0.0077$ & $0.119\pm0.021$ & $3.56^{+0.03}_{-0.04}$ \\
\hline
\multicolumn{8}{l}{$\boldsymbol{\mathrm{\alpha\mathchar`-\rm EDE}}$} \\
CMB+SH0ES & $70.90^{+0.80}_{-0.62}$ & $2.166^{+0.032}_{-0.023}$ & $0.9795\pm0.0038$ & $0.3047^{+0.0049}_{-0.0056}$ & $0.8372^{+0.0096}_{-0.0107}$ & $0.092^{+0.021}_{-0.018}$ & $3.51\pm0.04$ \\
CMB+DESI+SH0ES & $71.52\pm0.72$ & $2.179\pm0.026$ & $0.9829^{+0.0036}_{-0.0042}$ & $0.2973^{+0.0030}_{-0.0036}$ & $0.8261^{+0.0064}_{-0.0074}$ & $0.095^{+0.020}_{-0.017}$ & $3.51^{+0.04}_{-0.03}$ \\
CMB+JWST+SH0ES & $71.44^{+0.79}_{-0.72}$ & $2.183\pm0.027$ & $0.9833\pm0.0039$ & $0.3044\pm0.0050$ & $0.8430\pm0.0086$ & $0.108^{+0.019}_{-0.017}$ & $3.51^{+0.04}_{-0.03}$ \\
CMB+DESI+JWST+SH0ES & $72.18\pm0.84$ & $2.209\pm0.030$ & $0.9868\pm0.0044$ & $0.2972\pm0.0034$ & $0.8352^{+0.0078}_{-0.0068}$ & $0.113\pm0.022$ & $3.52^{+0.04}_{-0.03}$ \\
CMB+DESI+PPS & $72.04\pm0.76$ & $2.183\pm0.027$ & $0.9833\pm0.0041$ & $0.2980\pm0.0034$ & $0.8304\pm0.0072$ & $0.111^{+0.020}_{-0.018}$ & $3.51\pm0.03$ \\
\hline
\multicolumn{8}{l}{$\boldsymbol{\mathrm{RnR\mathchar`-EDE}}$} \\
CMB+SH0ES & $69.95^{+0.55}_{-0.40}$ & $2.148^{+0.022}_{-0.027}$ & $0.9815^{+0.0037}_{-0.0030}$ & $0.3084\pm0.0047$ & $0.8404\pm0.0086$ & $0.069^{+0.015}_{-0.012}$ & $3.49^{+0.06}_{-0.04}$ \\
CMB+DESI+SH0ES & $70.96^{+0.75}_{-0.56}$ & $2.177^{+0.026}_{-0.030}$ & $0.9869^{+0.0048}_{-0.0039}$ & $0.2993\pm0.0030$ & $0.8308\pm0.0080$ & $0.079^{+0.022}_{-0.015}$ & $3.50^{+0.05}_{-0.04}$ \\
CMB+JWST+SH0ES & $71.47\pm0.82$ & $2.181\pm0.031$ & $0.9900\pm0.0044$ & $0.3070\pm0.0056$ & $0.8522^{+0.0101}_{-0.0114}$ & $0.109\pm0.021$ & $3.49^{+0.04}_{-0.03}$ \\
CMB+DESI+JWST+SH0ES & $71.76^{+0.72}_{-0.80}$ & $2.204^{+0.026}_{-0.030}$ & $0.9923\pm0.0045$ & $0.2986\pm0.0033$ & $0.8393\pm0.0076$ & $0.100\pm0.019$ & $3.49^{+0.04}_{-0.03}$ \\
CMB+DESI+PPS & $71.41^{+0.69}_{-0.76}$ & $2.178\pm0.027$ & $0.9888\pm0.0044$ & $0.2995\pm0.0034$ & $0.8339\pm0.0076$ & $0.093\pm0.019$ & $3.50^{+0.05}_{-0.03}$ \\
\hline
\multicolumn{8}{l}{$\boldsymbol{\mathrm{AdS\mathchar`-EDE}}$} \\
CMB+SH0ES & $70.41\pm0.59$ & $2.136\pm0.025$ & $0.9835^{+0.0036}_{-0.0042}$ & $0.3036\pm0.0051$ & $0.8405\pm0.0092$ & $0.069^{+0.012}_{-0.014}$ & $3.42\pm0.01$ \\
CMB+DESI+SH0ES & $71.02^{+0.49}_{-0.73}$ & $2.150\pm0.025$ & $0.9857\pm0.0035$ & $0.2977\pm0.0035$ & $0.8336\pm0.0072$ & $0.073^{+0.012}_{-0.017}$ & $3.41\pm0.01$ \\
CMB+JWST+SH0ES & $70.94^{+0.76}_{-0.67}$ & $2.152\pm0.026$ & $0.9859\pm0.0036$ & $0.3032^{+0.0044}_{-0.0050}$ & $0.8478\pm0.0085$ & $0.082\pm0.014$ & $3.41\pm0.01$ \\
CMB+DESI+JWST+SH0ES & $71.43\pm0.62$ & $2.169\pm0.026$ & $0.9883\pm0.0035$ & $0.2978\pm0.0034$ & $0.8419\pm0.0074$ & $0.085\pm0.014$ & $3.40\pm0.01$ \\
CMB+DESI+PPS & $71.21\pm0.59$ & $2.148\pm0.025$ & $0.9856\pm0.0033$ & $0.2985\pm0.0032$ & $0.8368\pm0.0076$ & $0.080\pm0.014$ & $3.41\pm0.01$ \\
\hline
\end{tabular}
}
\end{table*}

{Table~\ref{table_qdmap} summarizes the residual $H_0$ tension quantified by $\Delta\chi^2_{H_0}$ and $Q_{\mathrm{DMAP}}$. In the $\Lambda$CDM model, the tension remains significant, with $Q_{\mathrm{DMAP}}>4\sigma$ for all considered combinations. Including the SH0ES calibration (PPS) provides a direct late-time anchor and allows EDE models to reduce the tension, while the inclusion of JWST data provides an independent probe of the early structure formation and leads to comparable or stronger improvements without relying on the local distance calibration.}

{For the full CMB+DESI+JWST combination, the axion-like EDE model achieves the strongest reduction of the tension, reaching $Q_{\mathrm{DMAP}}=0.6\sigma$, which is comparable to or even better than the corresponding constraints obtained with the SH0ES-calibrated dataset. The other EDE scenarios also show significant reductions, with the remaining tension generally below $2\sigma$. These results demonstrate that JWST observations provide complementary information to traditional distance-ladder measurements and play an important role in testing EDE scenarios by simultaneously constraining the expansion history and high-redshift structure formation.}

\begin{table*}[!htb]
\renewcommand\arraystretch{1.2}
\centering
\caption{{The $H_0$ tension quantified by the $\Delta\chi^2_{H_0}$ and $Q_{\mathrm{DMAP}}$ metrics for $\Lambda$CDM and the four EDE models.}}
\label{table_qdmap}
\setlength{\tabcolsep}{6pt}
\small
\begin{tabular}{lcc|cc|cc|cc|cc}
\hline\hline
& \multicolumn{2}{c|}{$\boldsymbol{\Lambda\mathrm{CDM}}$} & \multicolumn{2}{c|}{$\boldsymbol{\mathrm{Axion\mathchar`-EDE}}$} & \multicolumn{2}{c|}{$\boldsymbol{\mathrm{\alpha\mathchar`-EDE}}$} & \multicolumn{2}{c|}{$\boldsymbol{\mathrm{RnR\mathchar`-EDE}}$} & \multicolumn{2}{c}{$\boldsymbol{\mathrm{AdS\mathchar`-EDE}}$} \\
\cline{2-11}
Data & $\Delta\chi^2_{H_0}$ & $Q_{\mathrm{DMAP}}$ & $\Delta\chi^2_{H_0}$ & $Q_{\mathrm{DMAP}}$ & $\Delta\chi^2_{H_0}$ & $Q_{\mathrm{DMAP}}$ & $\Delta\chi^2_{H_0}$ & $Q_{\mathrm{DMAP}}$ & $\Delta\chi^2_{H_0}$ & $Q_{\mathrm{DMAP}}$ \\
\hline
CMB & $27.0$ & $5.2\sigma$ & $7.9$ & $2.8\sigma$ & $8.4$ & $2.9\sigma$ & $11.6$ & $3.4\sigma$ & $11.0$ & $3.3\sigma$ \\
CMB+DESI & $20.5$ & $4.5\sigma$ & $4.2$ & $2.1\sigma$ & $3.3$ & $1.8\sigma$ & $6.5$ & $2.6\sigma$ & $6.1$ & $2.5\sigma$ \\
CMB+JWST & $27.2$ & $5.2\sigma$ & $2.2$ & $1.5\sigma$ & $5.3$ & $2.3\sigma$ & $6.1$ & $2.5\sigma$ & $8.3$ & $2.9\sigma$ \\
CMB+DESI+JWST & $21.2$ & $4.6\sigma$ & $0.4$ & $0.6\sigma$ & $1.1$ & $1.0\sigma$ & $3.5$ & $1.9\sigma$ & $3.2$ & $1.8\sigma$ \\
CMB+DESI+PP & $20.6$ & $4.5\sigma$ & $3.2$ & $1.8\sigma$ & $3.4$ & $1.8\sigma$ & $7.8$ & $2.8\sigma$ & $7.0$ & $2.6\sigma$ \\
\hline
\end{tabular}
\end{table*}

\end{document}